\documentclass{article}
\usepackage{latexsym}
\usepackage[dvips]{graphicx}
\usepackage{comment}

\begin{document}

\newcommand{\be}{\begin{equation}}
\newcommand{\ee}{\end{equation}}
\newcommand{\bea}{\begin{eqnarray}}
\newcommand{\eea}{\end{eqnarray}}
\newcommand{\non}{\nonumber}
\newcommand{\nod}{\noindent}
\newcommand{\dpl}{\displaystyle \partial}
\newcommand{\ba}{\begin{array}}
\newcommand{\ea}{\end{array}}
\newcommand{\bc}{\begin{center}}
\newcommand{\ec}{\end{center}}
\newcommand{\und}{\underline}
\newcommand{\mb}{\mbox}
\newcommand{\ov}{\overline}

\title{\bf The structure of flame filaments in chaotic flows\\}

\author{I. Kiss$^1$, J.H. Merkin$^{1,\ast}$, S.K. Scott$^2$, P.L. Simon$^2$, S. Kalliadasis$^3$, Z. Neufeld$^4$}

\maketitle

\nod
$^1$  Department of Applied Mathematics\\
$^2$ Department of Chemistry\\
$^3$ Department of Chemical Engineering\\
$\quad$ University of Leeds\\
\quad Leeds, LS2 9JT, \ UK\\

\nod
$^4$ Department of Applied Mathematics and Theoretical Physics\\
University of Cambridge\\
Silver Street\\
Cambridge CB3 9EW, \ UK\\

\nod
$\ast$ corresponding author\\
email: amtjhm@amsta.leeds.ac.uk\\
FAX: +113 233 5090\\

\begin{abstract}

The structure of flame filaments resulting 
from chaotic mixing within a combustion reaction is considered. The transverse profile of the filaments is investigated numerically and
analytically based on a one-dimensional model that represents
the effect of stirring as a convergent flow.
The dependence of the steady solutions 
on the Damk\"ohler number and Lewis number is treated in detail. 
It is found that, below a critical Damk\"ohler number $Da_{crit}$, the flame
is quenched by the flow. The quenching transition appears as a result of
a saddle-node bifurcation where the stable steady filament solution
collides with an unstable one. The shape of the steady solutions for
the concentration and temperature profiles changes with the Lewis
number and the value of $Da_{crit}$ increases monotonically with 
the Lewis number. Properties of the solutions are studied analytically in the
limit of large Damk\"ohler number and for small and large Lewis number.\\

\end{abstract}

\nod

\nod

\newpage

\baselineskip=20pt

\section{Introduction}

Many chemical and biological processes take place within an imperfectly 
mixed environment, examples include the mixing of reactants within continuously fed or batch reactors \cite{Epstein} and in the spread of plankton blooms within oceanic currents \cite{Abraham,GRL}. In these situations the time--dependent fluid flow, within which the reactions are taking place, can lead to the chaotic transport of fluid elements \cite{Aref,Ottino1,Ottino2,Solomon}, which become stretched into thin elongated filaments. This behaviour has been observed both experimentally by inserting dye droplets into the flow \cite{Chaiken,Ottino3,Ottino4,Ottino5,Gollub} and in numerical simulations \cite{Ottino5,Alvarez}.\\

In a two-dimensional system, a convergent and a divergent direction can be assigned to any point in the flow associated with the eigenvectors corresponding to the negative and positive Lyapunov exponents $-\lambda$ and $\lambda$ of the chaotic advection \cite{Giona,Varosi}. These directions are, respectively, tangent to the stable and unstable foliations of the advection dynamics. Any advected material line tends to align along the unstable foliation in forward time, or along the stable foliation in backward time. Thus, the stirring process smoothes out the concentration of any advected tracer along the stretching direction, whilst enhancing concentration gradients in the convergent direction,  producing a quasi-one-dimensional `lamellar' structure. \\

This allows us to separate the original reaction-advection-diffusion problem along the (Lagrangian) stretching and converging directions. In the stretching direction any perturbation is spread by the advective transport. This is the dominant process in this case and is much faster than diffusion. In the convergent direction, however, all three processes of reaction, advection and diffusion are of equal importance and need to be considered together. Chemical reactions of the type
$A + B \to P$ taking place in one-dimensional lamellar systems were
first studied numerically by Muzzio and Ottino \cite{Muzzio1,Muzzio2,Muzzio3} 
and analytically by Sokolov and Blumen \cite{Sokolov}. More recently, 
Clifford et al. investigated the evolution of a two-step
(competitive-consecutive) chemical reaction in lamellar systems
\cite{Clifford1,Clifford2}.\\

For a spatially smooth velocity field the relative motion of 
nearby fluid elements can be approximated by linearising the velocity
field along the path of a fluid particle. Thus, in the convergent direction
fluid elements approach each other with a relative velocity proportional
to their separation, $\delta v \sim -a(t) \delta r$, where the rate of 
convergence generally fluctuates in time along the trajectory. 
The average behaviour
can be  approximated by replacing $a(t)$ with its long time average given
by $\lambda$, the Lyapunov exponent of the chaotic advection.
In the context of autocatalytic type reactions
it has been shown, that the resulting one-dimensional 
`Lagrangian filament model' can be used to describe the mean transverse
profiles of filaments that propagate along the divergent direction
following the unstable foliation \cite{excitablePRL,CHAOS}.\\

Fluid mixing has an important role in combustion \cite{Peters,
Constantin1,Vulpiani,Ryzhik,Constantin2}.
Our main aim is to consider the effect of chaotic mixing within 
a combustion reaction, which we model as a first--order process 
converting a fuel $C$ to an inert product $P$ through the reaction
\be
C \to P \qquad \mb{rate  =  }c\, k(T)
\ee
\nod
with exothermicity $q$. $T$ is the (absolute) temperature and $c$ is the concentration of reactant $C$. The temperature dependence of the reaction rate is given by an Arrhenius law with an ignition temperature $T_i$, namely
\be
k(T) = \left\{
\begin{array}{ll}
k_0 \, \mb{exp}(-\frac{E}{RT}), &  \quad \mbox{if } \ T > T_i\\
0, & \quad \mbox{if  } \ T\leq T_i
\end{array} \right.
\ee
\nod
where $E$ and $R$ are the activation energy and the universal gas constant and $k_0$ is the (constant) pre-exponential factor. Our reason for choosing this form for the rate law is to avoid the difficulties associated with the cold boundary problem. An alternative approach is, rather artificially, to set the ambient temperature to zero. The form given by (2) allows for a non-zero ambient temperature and the discontinuity in $k(T)$ does not lead to any problems provided $T_i$ is kept small.\\

Here we are concerned with the structures of filaments that arise within the chaotic mixing and so, for present purposes, the nature of the time--dependent fluid flow is not important. This aspect will be described in a future paper \cite{Kiss} along with details of the resulting chaotic transport and reaction processes. This leads us to consider the equations for the filament structure,
\be
\rho C_p\left(\frac{\dpl T}{\dpl t}-\lambda  x \frac{\dpl T}{\dpl x}\right)=\kappa\frac{\dpl^2 T}{\dpl x^2}+q\, c \,k(T)
\ee
\be
\frac{\dpl c}{\dpl t}-\lambda x \frac{\dpl c}{\dpl x}=D\frac{\dpl^2 c}{\dpl x^2}-c \, k(T)
\ee
\nod
on $-\infty<x<\infty, \, t>0$. Here $x$ measures distance transverse to the filament, $t$ is time and $\rho$ is the density, $C_p$ the specific heat, $\kappa$ and $D$ are the thermal conductivity and diffusion coefficient, respectively.\\

The convective terms in equations (3,4) can be interpreted as advection by a pure strain flow at a constant stretching rate $-\lambda$ ($\lambda>0$) along the convergent direction. We can associate $\lambda$ with the Lyapunov exponent of the chaotic advection. Equations similar to (3,4) were first proposed
by Ranz \cite{Ranz} and have been considered previously for somewhat simpler chemical systems \cite{Sokolov,Clifford1} including autocatalytic reactions \cite{CHAOS} and excitable media \cite{excitablePRL}, where their significance in determining the nature of the chaotic mixing is brought out. 
These previous studies considered only chemical systems with identical 
diffusivities for all species. 
An important new aspect investigated in this paper is the
possibility of different diffusivities for heat and reactant, characteristic 
to the combustion process.\\

We apply the boundary conditions
\be
T\to T_a, \quad c \to c_0 \qquad \mb{as} \ |x| \to \infty \quad (t>0)
\ee
\nod
where $T_a$ is the ambient temperature (we are assuming that $T_a<T_i$). For the time--dependent problem we initiate the reaction by applying a local temperature input (above $T_i$).\\

To make equations (3,4) dimensionless, we introduce the variables
\be
\overline{T}=\frac{T-T_a}{T_b-T_a}, \quad \overline{c}=\frac{c}{c_0}, \quad \overline{t}=\lambda \, t, \quad \overline{x}=x \, \sqrt{\frac{\lambda}{D}} 
\ee
\nod
where $T_b$ is the burnt temperature $T_b=T_a+\frac{\displaystyle q c_0}{\displaystyle \rho C_p}$. This leads to the equations, on dropping the bars for convenience,
\be
\frac{\dpl T}{\dpl t}- x\frac{\dpl T}{\dpl x}=Le\frac{\dpl^2 T}{\dpl x^2}+Da \, c \, K(T)
\ee
\be
\frac{\dpl c}{\dpl t}- x \frac{\dpl c}{\dpl x}=\frac{\dpl^2 c}{\dpl x^2}-Da \, c \, K(T)
\ee
\nod
where
$$Da=\frac{k_0}{\lambda}, \qquad Le=\frac{\kappa}{\rho C_p D}$$
\nod
are the Damk\"ohler and Lewis numbers, respectively. The temperture dependence of the reaction (2) becomes
\be
K(T) = \left\{
\begin{array}{ll}
 \mb{exp}\left(-\frac{1}{\epsilon((1-\beta)T+\beta)}\right), &  \quad \mbox{if } \ T > \ov{T_i}\\
0, & \quad \mbox{if  } \ T\leq \ov{T_i}
\end{array} \right.
\ee
\nod
where
$$\epsilon=\frac{RT_b}{E}, \qquad \beta=\frac{T_a}{T_b}, \qquad \ov{T_i}=\frac{T_i-T_a}{T_b-T_a}.$$
\nod
The boundary conditions to be applied are that
\be
T\to 0, \quad c \to 1, \qquad \mb{as} \ |x| \to \infty
\ee

The Damk\"ohler number $Da$ characterises the ratio between the advective and the chemical time-scales. Large $Da$ corresponds to slow stirring or equivalently fast chemical reaction and {\em vice versa}. This, together with the Lewis number $Le$, are our main bifurcation parameters.\\

We start by considering the possible steady state solutions to equations (7,8) satisfying boundary conditions (10). 

\section{Steady states}

Here we consider the steady equations
\bea
Le \, T''+x \,T'+Da \, c \, K(T)&=&0\non\\
\\
c''+x\, c'-Da \, c \, K(T)&=&0\non
\eea
\nod
where $K(T)$ is given by (9) and where primes denote differentiation with respect to $x$. We look for symmetric solutions, applying the boundary conditions
\be
T'(0)=c'(0)=0, \qquad T \to 0, \ c \to 1 \quad \mb{as} \ x \to \infty.
\ee
 
When $Le=1$ we can combine equations (11) to eliminate the reaction terms. Integrating the resulting equation and applying (12) gives $T+c\equiv 1$ and then
\be
 T''+x \,T'+Da \, (1-T) \, K(T)=0
\ee  
\nod
again subject to (12).\\

Equations (11), or (13), were solved numerically using a standard NAG library routine for integrating boundary--value problems. The outer boundary conditions were applied at a large value of $x$ allowing them to be satisfied with sufficient accuracy. This value depended on the width of the filament which, in turn, depended on both the Lewis and Damk\"ohler numbers. The numerical integrations were found to be insensitive to point where the outer conditions were taken. The results are presented in figure 1, where we plot $T$ profiles for $Le=1$ (figure 1a) and both $T$ and $c$ profiles for $Le=0.1$ (figure 1b) and $Le=10$ (figure 1c) for representative values of $Da$ ($\epsilon=1.0, \, \beta =0.1, \, \ov{T_i}=0.001$). In these figures the broken lines represent temporally unstable solutions (see below). As $Da$ is increased for the stable solutions (full lines) the width of the filament increases with the temperature reaching a constant value within the central part of the filamant. This central temperature quickly reaches a value of unity for $Le=1.0$, whereas, for $Le=10.0$, it slowly approaches this value from below as $Da$ increases. The reactant $C$ is fully consumed in this central region of the filament. For $Le=0.1$, the reaction zone is much thinner than for the other two cases. Much higher temperatures can be achieved within the filament, though these reduce to unity, now from above, as $Da$ is increased. In all cases the unstable solutions (broken lines) reduce in both magnitude and extent as $Da$ increases.\\

To see how the solutions change as a given parameter is varied, we need some measure of the solution. The profiles shown in figure 1 suggest that the integrated quantity
\be
I_T=\int_0^{\infty}T(x) \, dx
\ee
is a suitable measure. Note that, if we add equations (11), integrate the resulting equation and apply boundary conditions (12), we find that $\displaystyle I_T=\int_0^{\infty}T(x)dx=\int_0^{\infty}(1-c(x))dx$. Graphs of $I_T$ against $Da$ are shown in figure 2 (for $Le=0.1, \, 1, \, 10$). The figure shows that there is a critical value of the Damk\"ohler number, $Da_{crit}$, with a saddle--node bifurcation at $Da_{crit}$ giving two solutions for $Da> Da_{crit}$ and no solutions for $Da< Da_{crit}$,
apart from the trivial uniform solution, $C(x)=1, T(x)=0$, corresponding to 
the extinction of the flame. Thus, the transition at $Da_{crit}$ can be interpreted as a quenching transition when the burning of the fuel is not sufficiently fast to
compensate for the diluting effect of the stirring, which 
tends to reduce the temperature of the filament.
$I_T$ increases with $Da$ on the upper solution branch and decreases on the lower branch. This effect can also be seen in the profiles shown in figure 1. Note that the value of $Da_{crit}$ appears to increase with $Le$.
We also note, that the possibility of quenching a flame by a sufficiently
strong shear flow has been shown recently in a rigorous mathematical study
by Constantin et al \cite{Constantin2}. 

We can compute $Da_{crit}$ as a function of the other parameters through looking for a non-trivial solution to the homogeneous linear equations that are obtained by making a small perturbation to the solution of equations (11) (or (13)). This point will be made clearer when we discuss the stability of the steady states in the next section. A graph of  $Da_{crit}$ against $Le$ computed in this way is shown in figure 3 (for $\epsilon=1.0, \, \beta =0.1, \, \ov{T}_i=0.001$). The figure shows that $Da_{crit}$ approaches a finite value (approximately 8.5 for this case) as $Le \to \infty$ and approaches a non-zero value as $Le \to 0$.\\

The graphs shown in figures 1--3 show different forms of behaviour depending on whether $Le$ is small or large. There is also a distinctive structure of the reaction zone when $Da$ is large and it is this limit that we now consider.

\subsection{Solution for large Da}

Here we describe how the upper branch (stable) solutions behave for $Da$ large. The form of the graphs shown in figure 1 suggest that, for $Da \gg 1$, the profiles have two regions of constant temperature and concentration. There is a central (fully reacted) region where $T=T_c, \, c=0$ and an outer (unreacted region) where $T=0, \, c=1$. There then must be a reaction zone which joins these two regions and which smoothes out the discontinuities in temperature and concentration. To determine the structure of this reaction zone we assume that it is centred on $x=x_0(Da)$. We then put
\be
X=Da^{1/2}(x-x_0)
\ee
\nod
and look for a solution by expanding
\bea
T(X;Da)=T_0(X)+Da^{-1}T_1(X)+\cdots,\non\\  c(X;Da)=c_0(X)+Da^{-1}c_1(X)+\cdots,\\
 T_c(Da)=T_c^{(0)}+Da^{-1}T_c^{(1)}+\cdots\non
\eea
\nod
A consideration of the resulting equations suggests that $x_0$ is of $O(Da^{1/2})$. This suggests that we put 
\be
x_0(Da) = Da^{1/2}(a_0+a_1 Da^{-1}+\cdots), \qquad Da \gg 1
\ee
\nod
where the $a_i$ are constants to be determined.\\

The equations at leading order are then   
\bea
Le \, T_0''+a_0 \,T_0'+ \, c_0 \, K(T_0)&=&0\non\\
\\
c_0''+a_0\, c_0'- \, c_0 \, K(T_0)&=&0\non
\eea
\nod
subject to the matching conditions
\be
T_0 \to 0, \quad c_0 \to 1 \quad \mb{as } \ X\to \infty, \qquad T_0 \to T_c^{(0)}, \quad c_0 \to 0 \quad \mb{as } \ X\to -\infty
\ee
\nod
Primes now denote differentiation with respect to $X$. If we add equations (18), integrate and apply the boundary conditions as $X\to \infty$ we obtain the equation
\be
LeT_0'+a_0T_0+c_0'+a_0c_0=a_0
\ee
\nod
If we now let $X\to -\infty$ in equation (20), we find, on using (19), that
\be
T_c^{(0)}=1
\ee
This result shows, from (6), that the central temperature within a filament is the burnt temperature $T_b$ for large Damk\"ohler numbers.\\

It is the solution to equations (18a) and (20) subject to (19) and using (21) that determines $a_0$. This boundary-value problem was solved numerically in a way similar to that used for determining the general filament solutions. A graph of $a_0$ against $Le$ is shown in figure 4 for $\epsilon=1.0,\, \beta=0.1, \, \ov{T_i}=0.001$. The figure shows that $a_0$, the position of the reaction front for large $Da$, increases with $Le$, in line with the profiles shown in figure 1.\\

We now consider the equations at $O(Da^{-1})$. We find, as before, that we can add these equations to eliminate the reaction terms to obtain
\be
LeT_1''+a_0T_1'+(X+a_1)T_0'+c_1''+a_0c_1'+(X+a_1)c_0'=0
\ee  
\nod The boundary conditions to be applied are that
\be
T_1\to T_c^{(1)}, \  c_1\to 0 \ \mb{as } \ X\to -\infty, \qquad T_1\to 0, \ c_1\to 0 \ \mb{as } \ X\to \infty
\ee 
\nod
If we integrate equation (22) and apply boundary conditions (19) and (23) we obtain, on using (21),
\be
a_0T_c^{(1)}=\int_{-\infty}^{\infty}(1-c_0-T_0)dX
\ee
\nod
The integral in (24) can then be evaluated as $(1-Le)/a_0$ by integrating equation (20). Thus we have
\be
T_c^{(1)}=\frac{1-Le}{a_0^2}, \qquad \mb{with} \quad T_c\sim 1+Da^{-1}\left(\frac{1-Le}{a_0^2}\right) \cdots \quad \mb{as } \ Da\to \infty
\ee
\nod
Expression (25) gives $T_c^{(1)}=0$ when $Le=1$ and shows why, in the central region, the burnt temperature is approached from below for $Le>1$ and from above for $Le<1$.   
 
\subsection{Solution for small $Le$}

Here we obtain a solution to equations (11) valid for $Le\ll 1$. We assume that $Da$ is of $O(1)$ and start by making the transformation $\xi=Le^{-1/2} \, x$. This leads to the equations
\bea
 T''+\xi \,T'+ Da\, c \, K(T)&=&0\non\\
\\
c''+Le\left(\xi\, c'- Da\, c \, K(T)\right)&=&0\non
\eea
\nod
still subject to (12), where primes now denote differentiation with respect to $\xi$. Equations (26) suggest looking for a solution by expanding in powers of $Le$. However, we find that this is not sufficient and we need to expand in powers of $Le^{1/2}$, namely
\be
T=T_0+Le^{1/2}\,T_1+Le\,T_2+\cdots, \qquad c=c_0+Le^{1/2}\,c_1+Le\,c_2+\cdots
\ee

At leading order $c_0''=0$ from which it follows that
\be
c_0\equiv 1
\ee
\nod
and then
\be
 T_0''+\xi \,T_0'+ Da\,  K(T_0)=0
\ee
\nod
The numerical solution of equation (29), subject to (12), shows that it has a similar structure to the results shown in figures 1 and 2. From the co-ordinate transformation $I_T=Le^{1/2}\,\ov{I}_T$ where
 $\displaystyle \ov{I}_T=\int_0^{\infty} T_0(\xi)d\xi$. A graph of $\ov{I}_T$ against $Da$ is shown in figure 5. The main point to note about this figure is the existence of a critical Damk\"ohler number $Da_{crit}=3.0136$ with two solution branches for $Da> Da_{crit}$. This is the limiting value of $Da_{crit}$ at $Le=0$ in figure 3. Note that the value of $I_T$ at $Da_{crit}$ decreases with $Le$ with $I_T \sim 1.530Le^{1/2}$ as $Le \to 0$.\\

At $O(Le^{1/2})$ we have $c_1''=0$. Hence $c_1=b_1$, where $b_1$ is a constant to be determined. At $O(Le)$ we then have
\be
c_2''=Da\,K(T_0), \qquad c_2'(0)=0
\ee
\nod
We can integrate equation (30) and using the result obtained from equation (29) that\\
 $\displaystyle Da\int_0^{\infty}K(T_0)d\xi=\ov{I}_T$, we have that
\be
c_2\sim\ov{I}_T\xi+b_2 \quad \mb{as } \ \xi \to \infty
\ee
\nod
for some further constant $b_2$. A consideration of the equations for the $T_j, \, (j\ge 1)$ in expansion (27) shows that they can be solved so as to satisfy the outer bondary condition.\\

Expression (31) shows that we require an outer region in which $T\equiv 0$, in which $x$ is now the independent variable and $c=1+Le^{1/2} \, \ov{c}$. At leading order we obtain
\be
\ov{c}''+x\ov{c}'=0, \quad \ov{c}\sim b_1+\ov{I}_T\,x+\cdots \ \mb{as } \ x \to 0, \quad \ov{c} \to 0 \ \mb{as } \ x\to \infty
\ee
\nod
The required solution is
\be
\ov{c}=b_1\sqrt{\frac{2}{\pi}}\int_x^{\infty}\mb{exp}(-s^2/2)ds=b_1\left[1-\mb{erf}(x/\sqrt{2})\right]
\ee
\nod
Using the matching condition for $x$ small then gives $\displaystyle b_1=-\ov{I}_T\sqrt{\frac{\pi}{2}}$. The concentration at the centre of the filament is then
\be
c(0)=1-Le^{1/2}\,\ov{I}_T(Da)\sqrt{\frac{\pi}{2}}+\cdots, \qquad \mb{as } \ Le\to 0
\ee

We can now see why the terms of $Le^{1/2}$ are required in the expansion (27). Without these terms $b_1$ would effectively be zero, the solution to equation (30) at $O(Le)$ would be the same and again would not satisfy the outer boundary condition ((31) still holds). An outer region would still be needed and be given by (32) though with $b_1=0$. A solution to this problem cannot be obtained along the lines given by (33).\\

\subsection{Solution for large $Le$}

The structure of the solution for $Le \gg 1$ is illustrated in figure 6 for $Le=10^3$ and $Da=12.0$ ($\epsilon=1.0, \, \beta=0.1, \, \ov{T}_i=0.001$). The graph shows that there is an inner region around $x=0$ where $c$ is very small and $T$ is approximately a constant. We start our solution for $Le$ large in this inner region. We leave $x$ unscaled and look for a solution of equations (11) by expanding
\be
T(x;Le)=T_c+Le^{-(1+m)}\, T_1(x)\cdots, \qquad c(x;Le)=Le^{-m}c_1(x)+\cdots
\ee
\nod
where $m$ $(>0)$ and $T_c$ ($\neq 0$) are to be determined. The equation for $c_1$ is 
\be
c_1''+xc_1'-\alpha c_1=0, \qquad c_1'(0)=0
\ee
\nod
where $\alpha=DaK(T_c)$ is a constant. Equation (36) can be solved in terms of Confluent Hypergeometric Functions \cite{Slater} as
\be
c_1(x)=A_0 \, \mb{exp}(-x^2/2){}_1F_1(\frac{\alpha+1}{2};\frac{1}{2};\frac{x^2}{2})
\ee
\nod
for some constant $A_0$. From (37) it follows that
\be
c_1(x)\sim \frac{A_0\sqrt{\pi}}{\left(\frac{\alpha-1}{2}\right)!2^{\alpha/2}}\, x^\alpha \left(1+\cdots\right) \quad \mb{as } \ x \to \infty
\ee

The equation for $T_1(x)$ is $\displaystyle T_1''(x)=-\alpha c_1(x)$ which can be integrated to give
\be
T_1(x)=-\frac{\alpha A_0}{(\alpha+2)}\, \mb{exp}(-x^2/2){}_1F_1(\frac{\alpha+3}{2};\frac{1}{2};\frac{x^2}{2})+B_1
\ee
\nod 
for some further constant $B_1$. From (39) 
\be
T_1(x) \sim -\frac{A_0\alpha\sqrt{\pi}}{2(\alpha+2)\left(\frac{\alpha+1}{2}\right)!2^{\alpha/2}}\, x^{\alpha+2} \left(1+\cdots\right) \quad \mb{as } \ x \to \infty
\ee

An outer region is required to satisfy the outer boundary conditions. The form of equations (11) suggests that the appropriate scaling for this outer region is $\xi=Le^{-1/2} x$ with $c$ and $T$ both of $O(1)$. Applying this scaling in expressions (38) and (40) shows that we should take $\displaystyle m=\frac{\alpha}{2}$. With this, the problem for the outer region is, at leading order,
\be
T''+\xi T'+Da\, K(T)=0, \qquad \xi c'-Da\, K(T)=0
\ee
\nod
(primes denote differentiation with respect to $\xi$) subject to
\bea
c&\sim& \ov{A}_0\xi^\alpha+\cdots, \quad T\sim T_c-\frac{\alpha \ov{A}_0}{(\alpha+2)(\alpha+1)}\xi^{\alpha+2}+\cdots \ \mb{as } \ \xi \to 0,\non\\
c&\to& 1, \ T\to 0 \ \mb{as } \ \xi \to \infty
\eea
\nod
where $\displaystyle  \ov{A}_0=\frac{A_0\sqrt{\pi}}{2^{\alpha/2}\left(\frac{\alpha-1}{2}\right)!}$.\\

The problem given by (41,42) has to be solved numerically and it is this solution that determines $T_c$ (and $\ov{A}_0$). The results are shown in figure 7a, where we plot $\displaystyle \ov{I}_T=\int_0^{\infty}T(\xi) d\xi$ against $Da$ (for $\epsilon=1.0, \, \beta=0.1, \, \ov{T}_i=0.001$). Note that $I_T=Le^{1/2}\ov{I}_T$. The graph shows that there is a critical Damk\"ohler number $Da_{crit}=8.503$, with two solutions for $Da>Da_{crit}$. This value for $Da_{crit}$ is the limiting value for large $Le$ seen in figure 3. The numerical solution gives $T_c$ and hence the exponent $\displaystyle m = \frac{1}{2}Da \, \mb{exp}\left(-\frac{1}{\epsilon((1-\beta)T_c+\beta)}\right)$ for expansion (35) in the inner region. Graphs of $m$ and $T_c$ are shown in figure 7b.

\section{Stability of the steady states}

To determine the stability of the steady states, we make a perturbation of the form
\be
T(x,t)=T_s(x)+\mb{e}^{\omega t}T_1(x), \quad
c(x,t)=c_s(x)+\mb{e}^{\omega t}c_1(x)
\ee
\nod
where $T_s$ and $c_s$ are the steady states (solutions of equations (11,12)) and $T_1$ and $c_1$ are small. We must allow for the possibility that $\omega$ and $T_1, \, c_1$ could be complex. Substituting (43) into equations (7,8) and linearizing gives
\bea
LeT_1''+xT_1'+Da(c_1K(T_s)+T_1c_sK'(T_s))-\omega T_1&=&0\non\\
\\
c_1''+xc_1'-Da(c_1K(T_s)+T_1c_sK'(T_s))-\omega c_1=0\non
\eea
\nod
subject to
\be
T_1'(0)=c_1'(0)=0, \qquad T_1\to 0, \ c_1 \to 0 \ \mb{as } \ x \to \infty
\ee
\nod
Equations (44,45) are a homogeneous problem for the eigenvalue $\omega$ and to force a non-trivial solution in the numerical integrations we also apply the condition $T_1(0)=1$.\\

The turning points in the $I_T$ -- $Da$ curves (see figures 2,5,7) correspond to saddle--node bifurcations, i.e. to points where $\omega$ has a real zero. Thus solving equations (44,45) with $\omega=0$ (and regarding $Da$ as an unknown parameter) determines the value of $Da$ at the saddle--node bifurcation, i.e. determines $Da_{crit}$. This was the procedure that was used to determine the values of $Da_{crit}$ shown in figure 3. We can use this procedure to examine how $Da_{crit}$ varies with the other parameters.  We considered how $Da_{crit}$ varied with $\epsilon$. To do so we needed to think about the way in which we defined our Damk\"ohler number. Our form uses $k_0$ for the reaction rate, perhaps a more suitable form is to use the the reaction rate at the burnt gas temperature $T_b$ to define a Damk\"ohler number $\ov{Da}$, related to our choice by $\ov{Da}=\mb{e}^{-1/\epsilon} \, Da$. This change is not particularly significant for the larger values of $\epsilon$ that we used ($\epsilon \simeq 1$)  but can be significant for smaller values of $\epsilon$. In figure 8 we plot $\ov{Da}_{crit}$ against $\epsilon$ for $Le=1.0, \, \beta=0.1$ (and $\ov{T}_i=0.001$). The graph shows that we can have nontrivial solutions even for small values of $\epsilon$ though relatively large values of $\epsilon$ are required to have reasonable values for $\ov{Da}_{crit}$. For smaller values of $\epsilon$ ($\epsilon$ less than about $0.3$) the values of $\ov{Da}_{crit}$ increase very rapidly.\\ 

We solved equations (44,45) numerically for a range of values of $Le$, varying $Da$. In all cases we found that the lower solution branch was unstable, in fact on this branch $\omega$ was real with $\omega>0$. The upper branch solutions were found to be stable in all the cases considered. Generally we found $\omega$ to be real and negative, though for high values of $Le$ ($Le \sim 10^3$) we found that $\omega$ could become complex though always had $Re(\omega)<0$. We were unable to locate any parameter values at which the solution became unstable (and oscillatory) through a Hopf bifurcation at $Re(\omega)=0$. These conclusions were confirmed by numerical integrations of the initial-value problem (7,8,10). These showed that a threshold temperature input was necessary to generate a nontrivial solution and that, when this was applied, the corresponding upper branch solution was approached for $t$ large when $Da>Da_{crit}$. For $Da<Da_{crit}$ the numerical solution returned to the initial state and the reaction ceased as $t$ increased for all inputs.

\section{Conclusion}

The main feature to emerge from our discussion is the existence of a critical Damk\"ohler number $Da_{crit}$. An important aspect of our kinetic model (1,2) in the reaction-diffusion context is that any localized temperature input has to be above some threshold value if spatial or statiotemporal structures are to be sustained at large times. The effect of the mixing by the flow, seen in our model (3,4) through the steady converging flow $\lambda x$, is to remove the heat from where it is generated by the exothermic reaction. This has the effect of increasing the threshold input needed for steady structures to form at large times. Thus, for small values of the Damk\"ohler number, mixing is too strong, the heat generated by the reaction is dissipated, the reaction cannot be sustained and the system returns to its original state. This means that relatively large values of the Damk\"ohler number are required for nontrivial states to form within the filaments. This appears at a critical value $Da_{crit}$ of the Damk\"ohler number where there is a jump from the totally quenched state ($Da<Da_{crit}$) to a fully developed state ($Da>Da_{crit}$). Thus the values of $Da_{crit}$ found in the present study are an important guide in determining where sustained combustion will arise in the more general flow problem \cite{Kiss}.\\

We found that $Da_{crit}$ is not particularly sensitive to the Lewis number (for the values of the other parameters considered), see figure 3, increasing from about $3$ if $Le$ is put to zero and approaching a limit of about $8.5$ for large values of $Le$. However, the resulting temperature and concentration profiles are strongly dependent on both $Da$ and $Le$. For large values of $Da$ a central, fully reacted core develops which is separated from the outer conditions by a relatively thin reaction zone. In this case combustion effects are spread out over a wide area of the filament. A similar situation arises for large values of the Lewis number. Here the influence of heat conduction is very much stronger than the diffusion of the reactant. The effect is to increase the spread of the reaction region to $O(Le^{1/2})$, with a thinner inner region of constant temperature $T_c$ somewhat less than the burnt gas temperature, see figures 6,7b. For small values of the Lewis number the reaction region becomes much thinner, large temperatures can be achieved (relative to the burnt gas temperature) and only small amounts of fuel are consumed.\\

Criticality is an inherent part of the system and appears to be present for all values of the parameters associated with the Arrhenius kinetics. To get `reasonable' values for $Da_{crit}$ requires that $\epsilon$ be relatively large (figure 8), which can be regarded as corresponding to smaller activation energies. This is in contrast to other combustion problems where criticality is seen only for the higher activation energies and disappears (sometimes through a hysteresis bifurcation) at a given activation energy. Finally we note that we found only stable steady structures when $Da>Da_{crit}$. We found no further bifurcations to oscillatory behaviour.\\            

Finally, we note that the simple one-dimensional model (3,4) considered here is
a somewhat idealised model of the full advection-reaction-diffusion problem in which the effects of fluctuations of the stretching rate, curvature of the filaments,
interactions between filaments etc. are neglected. Such effects, however, may
play an important role in certain situations and deserve further investigations.

\

\nod
{\bf Acknowledgements}\\
We wish to acknowledge the support of the ESF Programme REACTOR and IK wishes to thank ORS and the University of Leeds for financial support.\\

\

\bc
{\bf References}
\ec

\newpage

\bc 
{\bf Captions for figures}\\
\ec

\nod
{\bf Figure 1.} \ (a) Temperature $T$ profiles for $Le=1$ ($Da$ takes values from $6.965$ to $100$ for the stable solutions and from $7$ to $75$ for the unstable solutions). Temperature $T$ and concentration $c$ profiles for (b) $Le=0.1$ ($Da$ takes values from $5$ to $70$ for the stable solutions and from $6$ to $20$ for the unstable solutions) and (c) $Le=10$ ($Da$ takes values from $8.5$ to $50$ for the stable solutions and from $8.7$ to $50$ for the unstable solutions), ($\epsilon=1.0, \, \beta=0.1, \, \ov{T}_i=0.001$). The broken lines represent temporally unstable solutions.\\

\nod
{\bf Figure 2.} \ Graphs of $I_T$ (defined by (14)) against $Da$ for $Le=0.1, \,1, \, 10$ ($\epsilon=1.0, \, \beta=0.1, \, \ov{T}_i=0.001$). The lower branch solutions are temporally unstable.\\

\nod
{\bf Figure 3.} \ A plot of $Da_{crit}$ against $Le$ ($\epsilon=1.0, \, \beta=0.1, \, \ov{T}_i=0.001$).\\

\nod
{\bf Figure 4.} \ The solution for $Da \gg 1$; $a_0$, the thickness of the central, fully reacted region, plotted against $Le$, ($\epsilon=1.0, \, \beta=0.1, \, \ov{T}_i=0.001$).\\ 

\nod
{\bf Figure 5.} \ The solution for $Le \ll 1$; $\ov{I}_T$, obtained from the numerical solution of equation (29), plotted against $Da$ ($\epsilon=1.0, \, \beta=0.1, \, \ov{T}_i=0.001$).\\ 

\nod
{\bf Figure 6.} \ The solution for large $Le$; concentration $c$ and temperature $T$ profiles for $Le=10^3$, $Da=12.0$ ($\epsilon=1.0, \, \beta=0.1, \, \ov{T}_i=0.001$).\\ 
 
\nod
{\bf Figure 7.} \ The solution for large $Le$; (a) $\ov{I}_T$, (b) the central temperature $T_c$ (full line) and the exponent $m$ (broken line) for the inner region expansion (35) plotted against $Da$,  ($\epsilon=1.0, \, \beta=0.1, \, \ov{T}_i=0.001$).\\ 

\nod
{\bf Figure 8.} \ A plot of $\ov{Da}_{crit}$ against $\epsilon$ ($\ov{Da}_{crit}= \mb{e}^{-1/\epsilon} \, Da_{crit}$) for $Le=1.0, \, \beta=0.1$  ($\ov{T}_i=0.001$).\\

\newpage
\begin{figure}[!hbp]
\centering
\includegraphics[width=14cm,height=25cm,angle=-90]{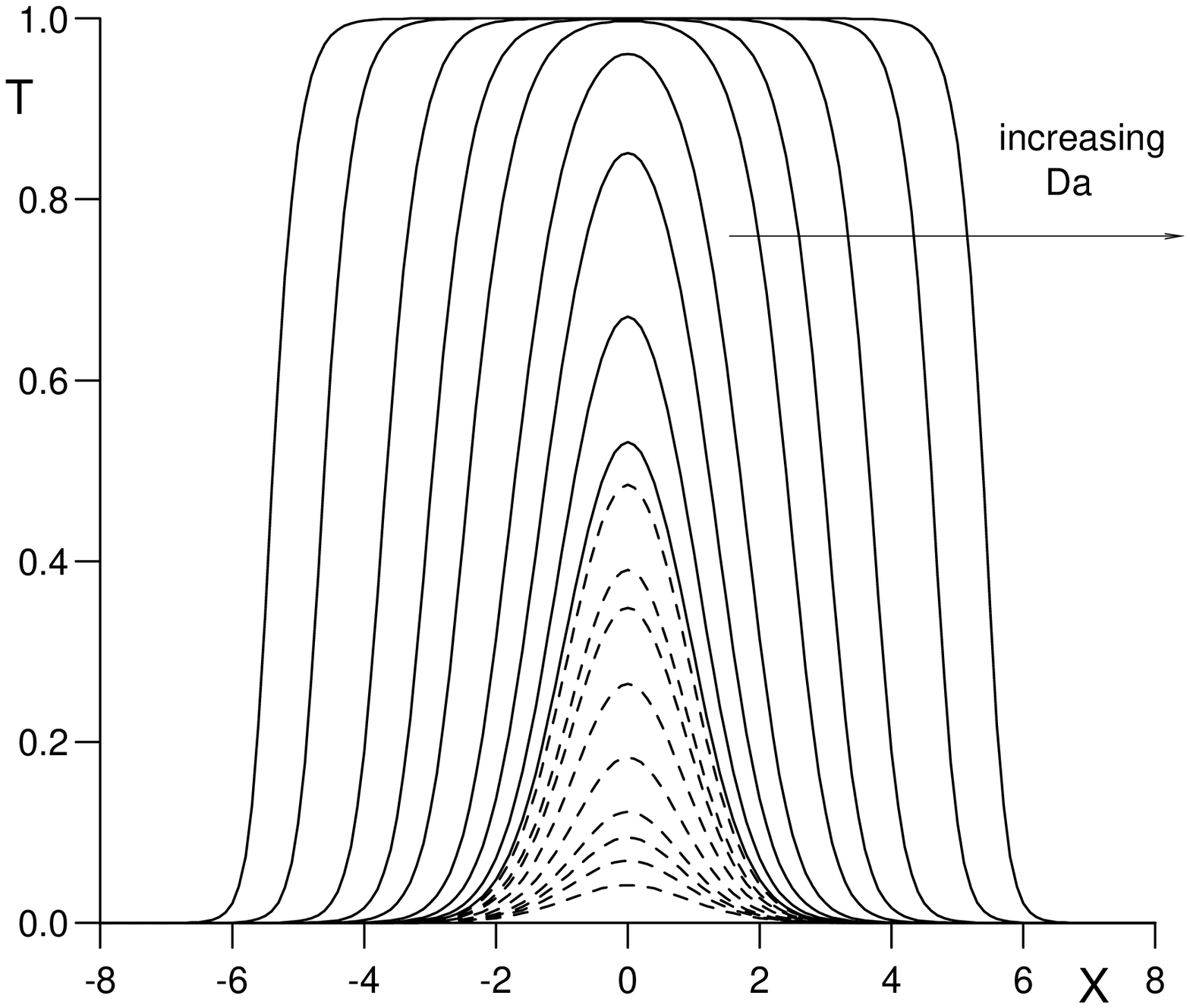}
\caption{Fig1aTemperature}
\end{figure}

\newpage
\begin{figure}[!hbp]
\centering
\includegraphics[width=14cm,height=25cm,angle=-90]{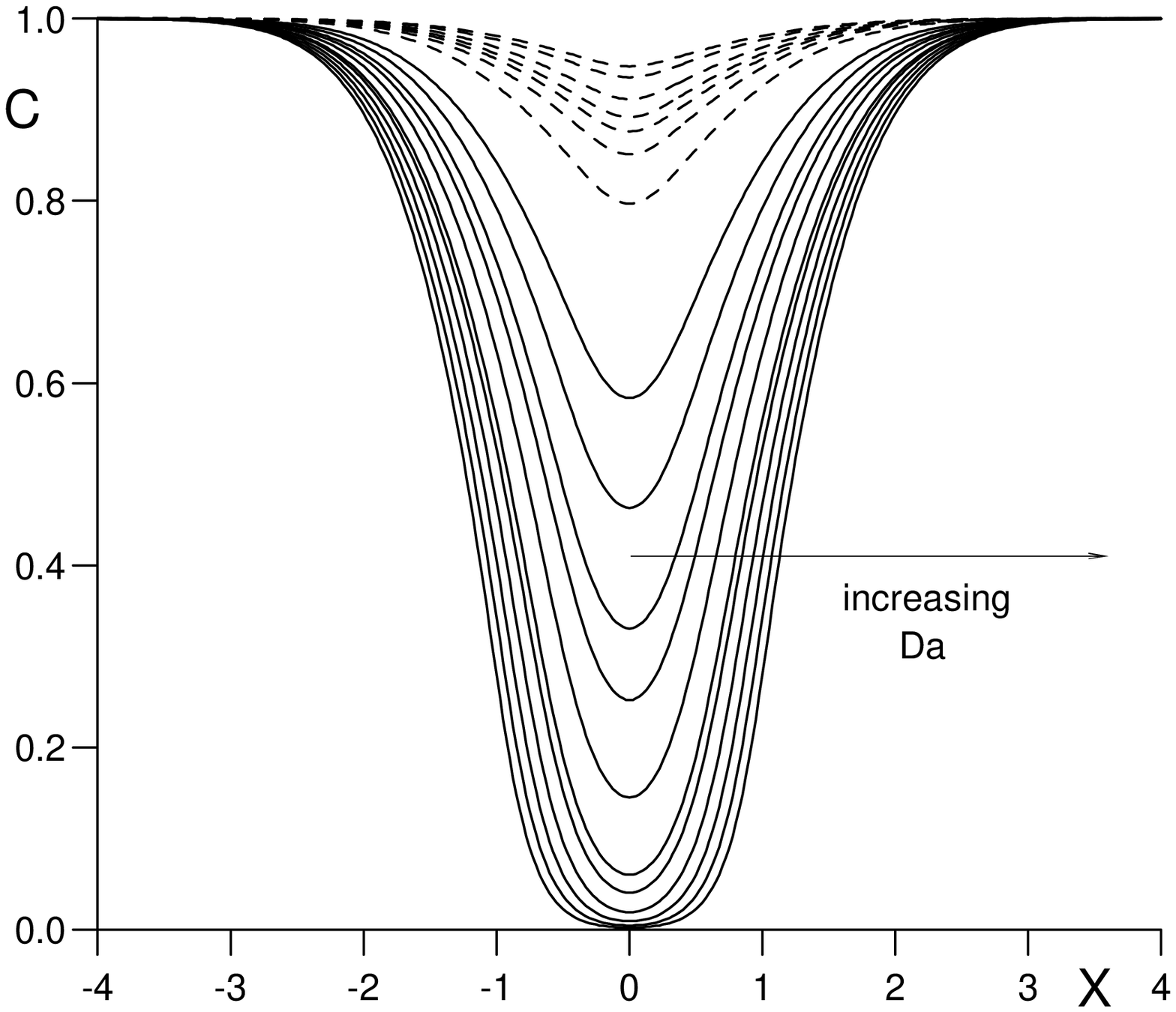}
\caption{Fig1bFuel}
\end{figure}

\newpage
\begin{figure}[!hbp]
\centering
\includegraphics[width=14cm,height=25cm,angle=-90]{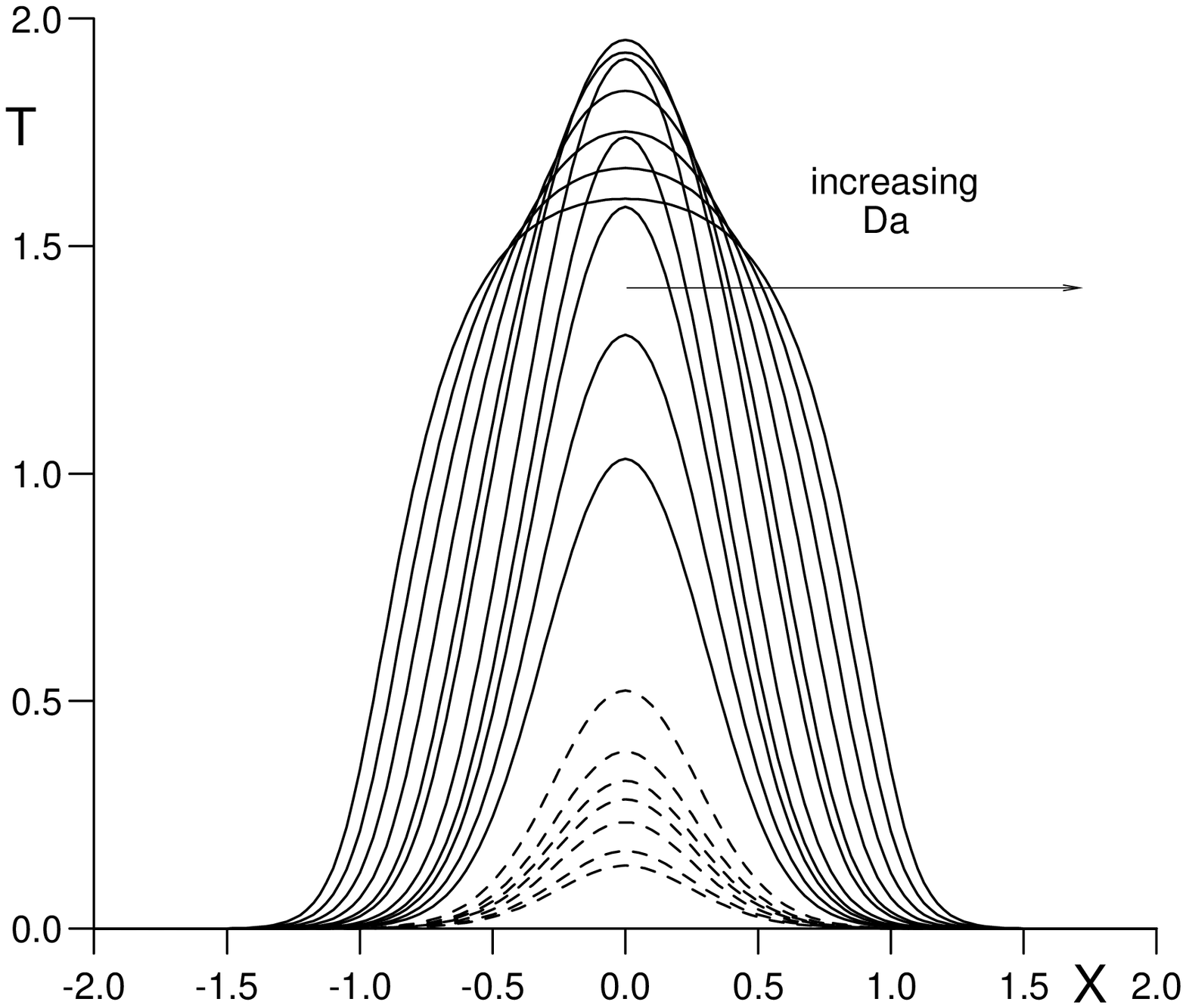}
\caption{Fig1bTemperature}
\end{figure}

\newpage
\begin{figure}[!hbp]
\centering
\includegraphics[width=14cm,height=25cm,angle=-90]{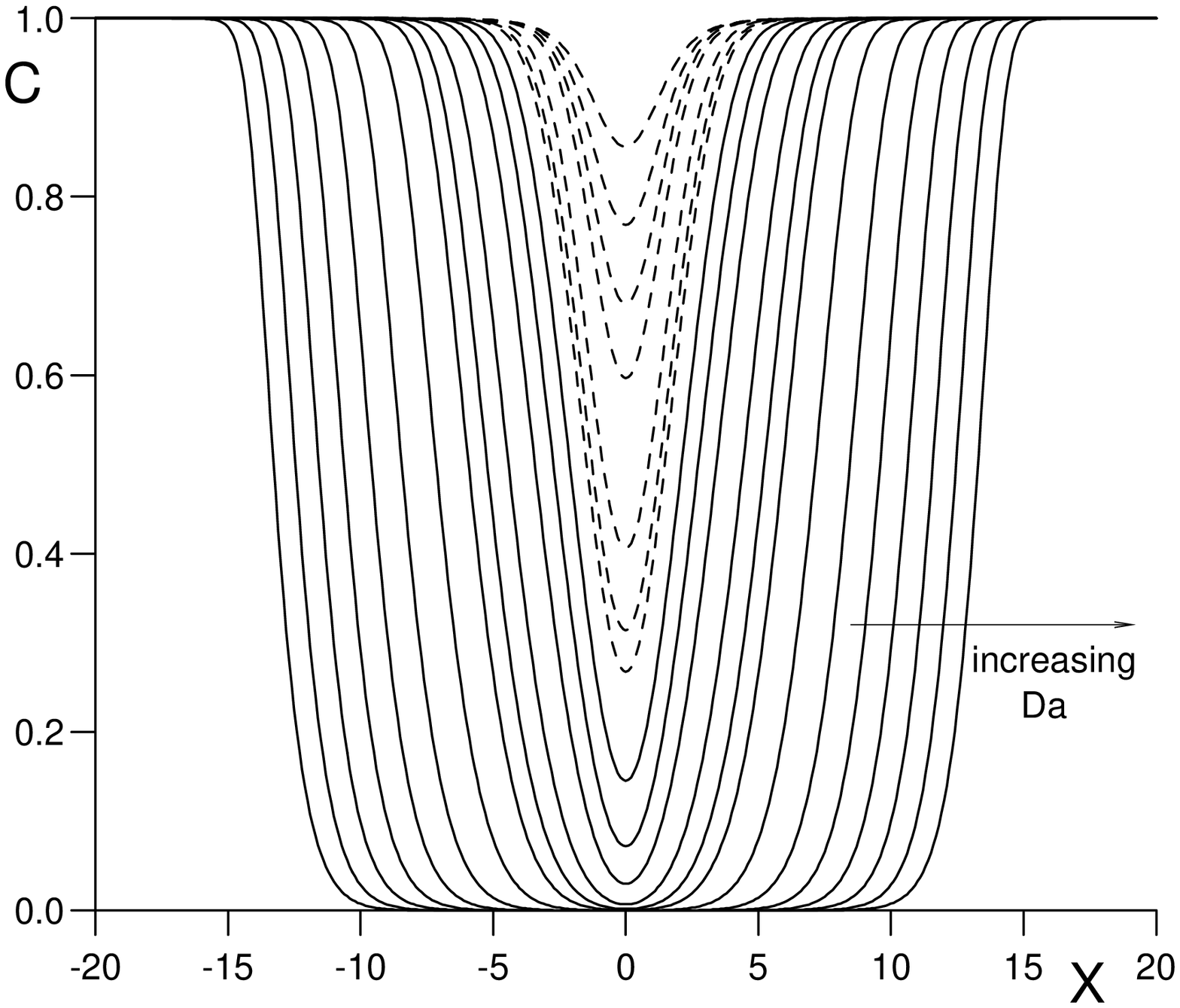}
\caption{Fig1cFuel}
\end{figure}

\newpage
\begin{figure}[!hbp]
\centering
\includegraphics[width=14cm,height=25cm,angle=-90]{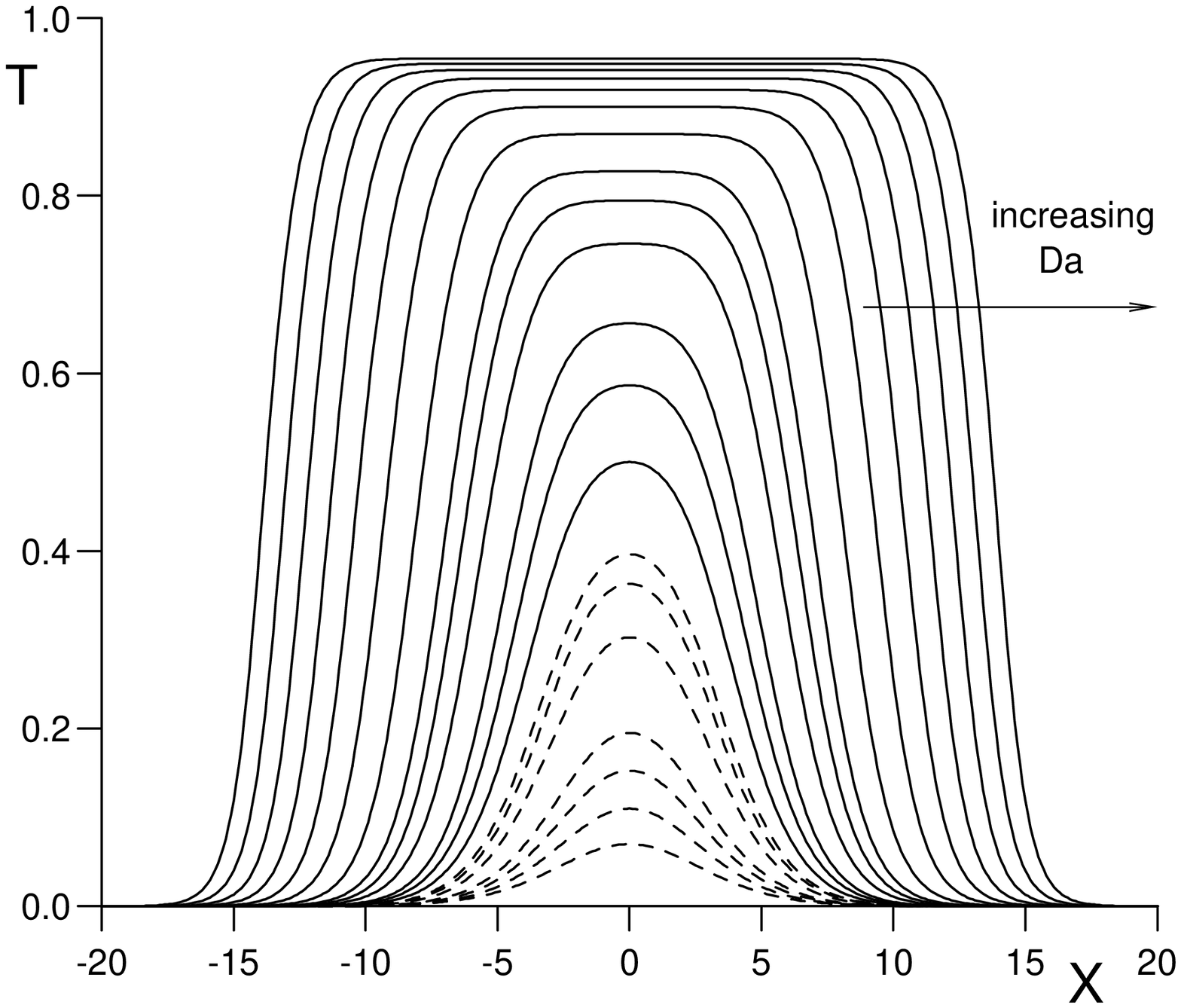}
\caption{Fig1cTemperature}
\end{figure}

\newpage
\begin{figure}[!hbp]
\centering
\includegraphics[width=14cm,height=25cm,angle=-90]{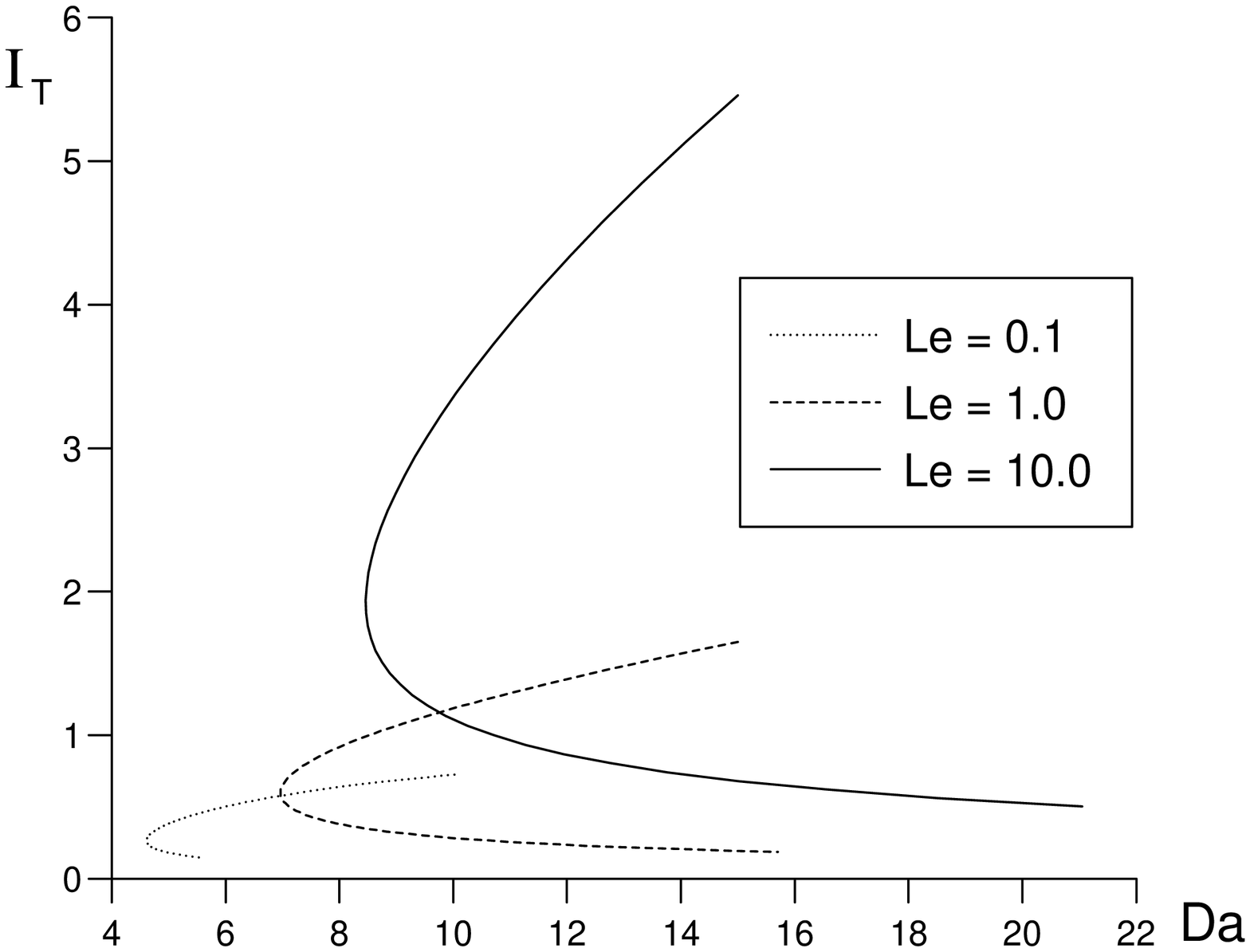}
\caption{Fig2}
\end{figure}

\newpage
\begin{figure}[!hbp]
\centering
\includegraphics[width=14cm,height=25cm,angle=-90]{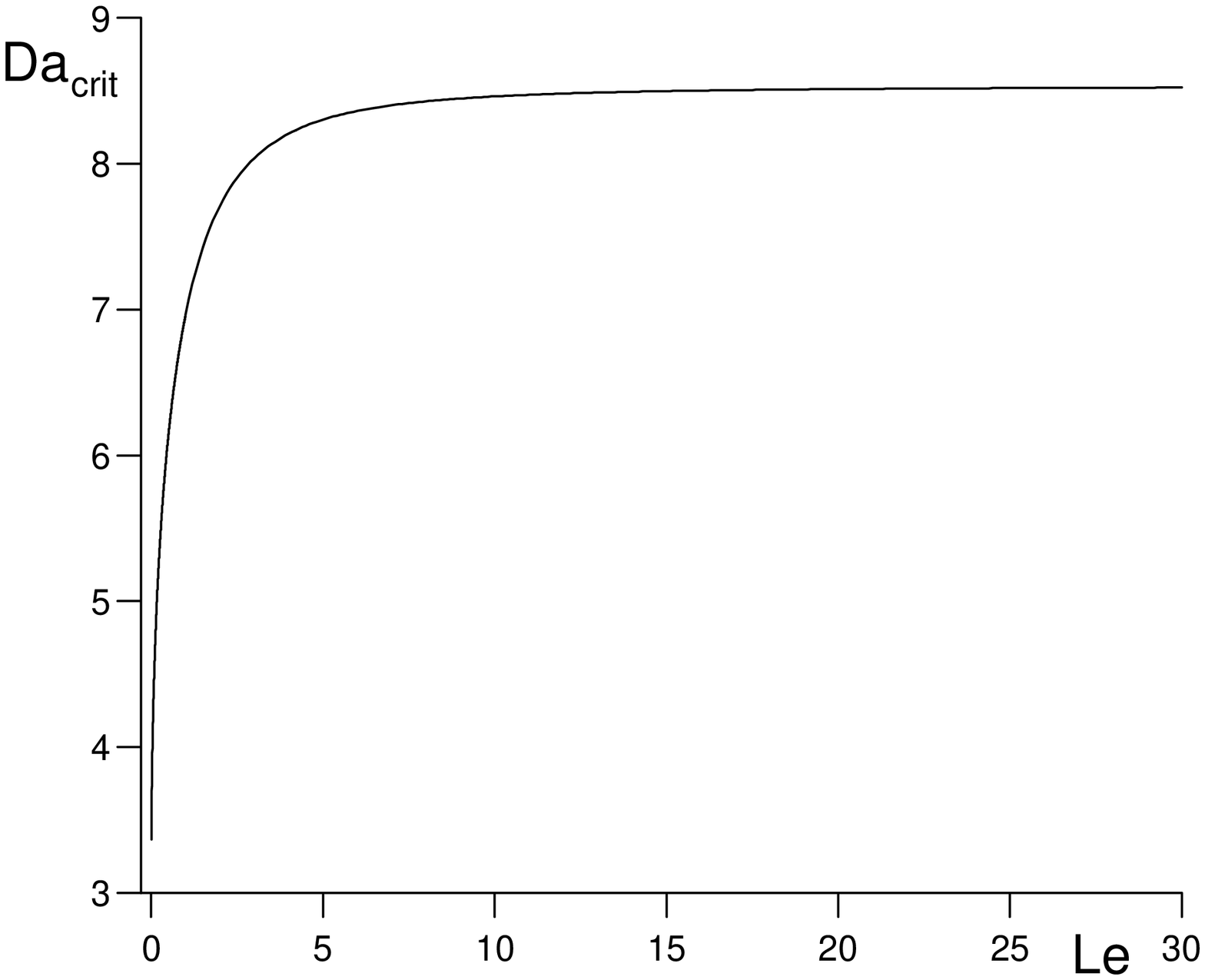}
\caption{Fig3}
\end{figure}

\newpage
\begin{figure}[!hbp]
\centering
\includegraphics[width=14cm,height=25cm,angle=-90]{fig4.plt}
\caption{fig4}
\end{figure}

\newpage
\begin{figure}[!hbp]
\centering
\includegraphics[width=14cm,height=25cm,angle=-90]{fig5.plt}
\caption{fig5}
\end{figure}

\newpage
\begin{figure}[!hbp]
\centering
\includegraphics[width=14cm,height=25cm,angle=-90]{fig6.plt}
\caption{fig6}
\end{figure}

\newpage
\begin{figure}[!hbp]
\centering
\includegraphics[width=14cm,height=25cm,angle=-90]{fig7a.plt}
\caption{fig7a}
\end{figure}

\newpage
\begin{figure}[!hbp]
\centering
\includegraphics[width=14cm,height=25cm,angle=-90]{fig7b.plt}
\caption{fig7b}
\end{figure}

\newpage
\begin{figure}[!hbp]
\centering
\includegraphics[width=14cm,height=25cm,angle=-90]{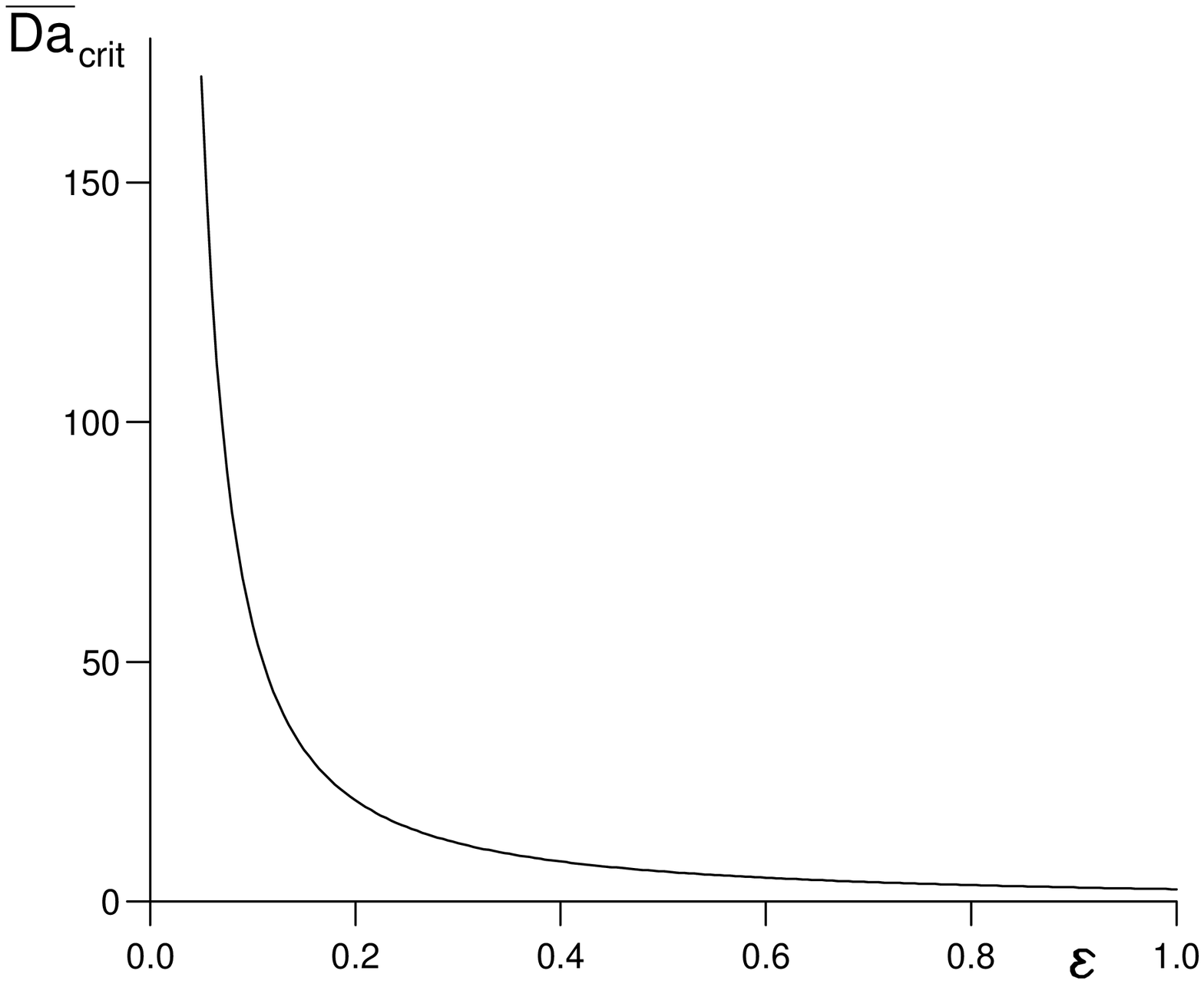}
\caption{Fig8}
\end{figure}
\end{document}